\documentclass[useAMS,usenatbib]{mn2e}

\usepackage{epsfig} 
\usepackage{multirow}
\usepackage{color}
\usepackage{graphicx}
\usepackage{amsmath}
\usepackage{amssymb}
\usepackage{booktabs}
\usepackage{float}

\newcommand{\be}{\begin{eqnarray}}
\newcommand{\ee}{\end{eqnarray}}
\newcommand{\bi}{\begin{itemize}}
\newcommand{\ei}{\end{itemize}}

\newcommand{\bs}{\boldsymbol}

\newcommand{\vk}{\vec{k}}

\newcommand{\vq}{\vec{q}}
\newcommand{\fnl}{f_{\rm NL}}
\newcommand{\gnl}{g_{\rm NL}}
\newcommand{\tnl}{\tau_{\rm NL}}
\newcommand{\nt}{\bar{n}_{\rm tot}}
\newcommand{\AVE}[1]{\langle#1\rangle}

\title[Testing Multi-Field Inflation with Galaxy Bias]
{Testing Multi-field Inflation with Galaxy Bias}

\author[Matteo Biagetti, Vincent Desjacques and  Antonio Riotto]
{Matteo Biagetti\thanks{Matteo.Biagetti@unige.ch}, 
Vincent Desjacques\thanks{Vincent.Desjacques@unige.ch}  and  
Antonio Riotto\thanks{Antonio.Riotto@unige.ch} \\
\\
D\'epartement de Physique Th\'eorique and Center for Astroparticle Physics (CAP)\\
Universit\'e de Gen\`eve, 24 quai Ernest Ansermet, CH-1211 Gen\`eve, Switzerland
}

\begin{document}

\pagerange{\pageref{firstpage}--\pageref{lastpage}}

\maketitle

\label{firstpage} 
\begin{abstract}
Multi-field models of inflation predict an inequality between the amplitude 
$\tnl$ of the collapsed limit of the four-point correlator of the primordial 
curvature perturbation and the amplitude $\fnl$ of the squeezed limit of its 
three-point correlator. 
While a convincing detection of non-Gaussianity through the squeezed limit 
of the three-point correlator would rule out all single-field models, a 
robust confirmation or disproval of the inequality between $\tnl$ and $\fnl$ 
would provide crucial information about the validity of multi-field models 
of inflation. In this paper, we discuss to which extent future measurements 
of the scale-dependence of galaxy bias can test multi-field inflationary 
scenarios. 
The strong degeneracy between the effect of a non-vanishing $\fnl$ and $\tnl$
on halo bias can be broken by considering multiple tracer populations of the 
same surveyed volume. If halos down to $10^{13}M_\odot/h$ are resolved in a 
survey of volume $25$(Gpc$/h)^3$, then testing multi-field models of inflation 
at the 3-$\sigma$ level would require, for instance, a detection of $\tnl$ at 
the level of $\tnl\sim 10^5$ given a measurement of a local bispectrum with 
amplitude $\fnl\sim 10$. However, we find that disproving multi-field models 
of inflation with measurements of the non-Gaussian bias only will be very 
challenging, unless $|\fnl|\gtrsim 80$ and one can achieve a halo mass 
resolution of $\sim 10^{10}M_\odot/h$.

\end{abstract}
\begin{keywords}
	cosmology: theory -- large scale structure of the universe -- inflation
\end{keywords}

\section{Introduction}

Inflation (see \cite{lrreview} for a review) has  become the dominant  paradigm 
for understanding the initial conditions for the large scale structure (LSS)  
formation and for Cosmic Microwave Background anisotropy (CMB). 
In the inflationary picture, primordial densities are created from quantum 
fluctuations ``redshifted'' out of the horizon during an early period of superluminal 
expansion of the universe, where they are ``frozen''. 
Perturbations at the surface of last scattering are observable as temperature 
anisotropy in the CMB. 
The last and most impressive confirmation of the inflationary paradigm has 
been recently provided by the data 
of the Wilkinson Microwave Anisotropy Probe (WMAP) mission which has 
marked the beginning of the precision era of the CMB measurements in space
(\cite{wmap7}).

Despite the simplicity of the inflationary paradigm, the mechanism by which  the 
cosmological curvature perturbation is generated  is not yet fully established. 
In the single-field models of inflation, the observed density perturbations are 
induced by fluctuations of the inflaton field itself. 
An alternative to the standard scenario is represented by the curvaton mechanism
(\cite{curvaton1}, \cite{LW}, \cite{curvaton3})  where the final curvature 
perturbations are produced from an initial isocurvature perturbation associated 
to the quantum fluctuations of a light scalar field (other than the inflaton), 
the curvaton, whose energy density is negligible during inflation. 
The curvaton isocurvature perturbations are transformed into adiabatic ones when 
the curvaton decays into radiation much after the end of inflation. 
Alternatives to the curvaton model are those models characterised by the curvature 
perturbation being generated by an inhomogeneity in the decay rate (\cite{rate1}, 
\cite{rate2})  of the particles responsible for the 
reheating after inflation.
Other opportunities for generating the curvature perturbation occur at the end 
of inflation (\cite{end1}, \cite{end2}) and during preheating (\cite{during}). 
A precise measurement of the spectral index $n_\zeta$ of the comoving curvature 
perturbation $\zeta$ will provide a powerful constraint to single-field models of  
inflation which predict  the spectral index to be  close to unity. 
However, alternative mechanisms, like the curvaton, also predict a value of the 
spectral index  very close to unity. Thus, even a precise measurement of the 
spectral index will not allow us to efficiently distinguish among them.
Furthermore, the lack of a gravity-wave signal in CMB anisotropies would not give 
us any information about the perturbation generation mechanism, since alternative 
mechanisms predict an amplitude of gravity waves far too small to be detectable by 
future experiments aimed at observing the $B$-mode of the CMB polarisation. 

There is, however, a third observable which will prove fundamental in providing 
information about the mechanism chosen by Nature to produce the structures we see 
today. It is the deviation from a Gaussian statistics, {\it i.e.}, the presence of 
higher-order connected correlation functions of the perturbations. 
Indeed, a possible source of non-Gaussianity (NG) could be primordial in origin, 
being specific to a particular mechanism for the generation of the cosmological 
perturbations (for a review see \cite{BKMR}). This is what makes a positive 
detection of NG so relevant: it might help discriminating among competing 
scenarios which, otherwise, would might remain indistinguishable.

To characterise the level of NG in the comoving curvature perturbation, one usually  
introduces two nonlinear parameters,  $\fnl$ and $\tnl$. The first one is defined  
in terms of the three-point correlator, the bispectrum, of the comoving curvature 
perturbation in the so-called squeezed limit
\be
\fnl=\frac{5}{12}\frac{\langle \zeta_{\vk_1}\zeta_{\vk_2}\zeta_{\vk_3}\rangle^\prime}
{P^\zeta_{\vk_1}P^\zeta_{\vk_2}}\,\,\,\,\,\,\,\,(k_1\ll k_2\sim k_3)~.
\ee
The second one is defined  in terms of the four-point correlator, the trispectrum,   
in the so-called collapsed limit
\be
\tnl=\frac{1}{4}\frac{\langle \zeta_{\vk_1}\zeta_{\vk_2}\zeta_{\vk_3}\zeta_{\vk_4}\rangle^\prime}
{P^\zeta_{\vk_1}P^\zeta_{\vk_3}P^\zeta_{\vk_{12}}}\,\,\,\,\,\,\,\,(\vk_{12}\simeq  0)~. 
\label{tf2}
\ee
We have normalised the correlators with respect to the power spectrum of the 
curvature perturbation, 
\be
\langle\zeta_{\vec{k}_{1}}\zeta_{\vec{k}_{2}}\rangle = 
(2\pi)^3\delta({\vec{k}_{1}}+{\vec{k}_{2}})P^\zeta_{\vec{k}_{1}}
\ee
and used the notation  $\vec{k}_{ij}=(\vec{k}_{i}+\vec{k}_{j})$. In all single-field 
models of inflation the bispectrum is suppressed in the squeezed limit and is non 
vanishing only when the spectral index deviates from unity, $\fnl=5/12(1-n_\zeta)\simeq 0.02$ 
(see \cite{acqua}, \cite{con1},\cite{con2}, \cite{con3}). 
A convincing detection of NG in the squeezed limit, $\fnl\gg 1$, would therefore rule 
out all single-field models (one should be aware though that, in single-field models of
inflation, a large NG can be generated in shapes others than the squeezed, {\it e.g.} 
in the equilateral configuration). However, such a detection would not rule out 
multi-field models of inflation where the NG is seeded by light fields other than the 
inflaton. 
How can  we derive some useful informations about them? 
In this respect, the collapsed limit of the four-point correlator is particularly 
important because, together with the squeezed limit of the three-point correlator, 
it may lead to the so-called Suyama-Yamaguchi (SY) inequality (\cite{SY}, see also 
\cite{SY1}, \cite{SY2}). 
Based on  the  conditions  that 1) scalar fields are responsible for generating curvature
perturbations and  that 2)  the fluctuations in the  scalar fields at the horizon crossing
are scale invariant and Gaussian, Suyama and Yamaguchi proved the inequality
\be
\label{SYin}
\tnl\ge \left(\frac{6}{5}\fnl\right)^2.
\ee
The  condition 2) amounts to assuming that the connected three- and four-point 
correlations of the light  fields vanish and that the NG is generated at super-horizon 
scales. This is quite a restrictive assumption. However, based on the operator product 
expansion, which is particularly powerful in characterising in their full generality 
the squeezed limit of the three-point correlator and the collapsed limit of the
four-point correlator, it was shown that the SY inequality holds also for NG light 
fields (\cite{KR}). 
This  is consequence of fundamental physical principles (like positivity of the 
two-point function) and its hard violation would require some new non-trivial physics 
to be involved.

The observation of a strong  violation of the inequality will then have profound 
implications for inflationary models. It will imply either that  multi-field 
inflation cannot be responsible for generating the observed fluctuations independently 
of the details of the model, or that some new non-trivial (ghost-like) degrees of freedom 
play a role during inflation (\cite{KR}).

Testing the SY inequality with future LSS observations and, therefore, the validity of 
multi-field inflationary models is the subject of this paper. The squeezed limit of 
the bispectrum and the collapsed limit of the trispectrum are  particularly interesting 
from the observationally point of view because they are associated to pronounced effects 
of NG on the clustering of dark matter halos and, in particular, to a strongly 
scale-dependent bias (\cite{dalal}). 
Measurements of the galaxy power spectrum have been exploited to set limits on primordial 
non-Gaussianity competitive with those inferred from CMB observations (\cite{slosar}, 
\cite{DS10}, \cite{X11}). 
As we have seen, a large value of $\fnl$ in the squeezed limit implies that the cosmological 
perturbations are generated within a multi-field model of inflation where the NG is sourced
 by light fields other than the inflaton. 
An inescapable consequence of the SY inequality (\ref{SYin}) is that the NG is also 
characterised by a large trispectrum in the collapsed limit. 
Therefore, investigations that take advantage of the scale-dependent effects of NG on the 
clustering of dark matter halos should in principle take into account both $\fnl$ and 
$\tnl$. However, since the contribution from the latter is suppressed by $10^{-4}(\tnl/\fnl)$,  
setting limits on $\fnl$ under the assumption $\tnl=0$, as done in the literature, should 
be a good approximation unless $|\tnl|\gg \fnl^2$.

In this paper, we will essentially try to answer the following question: what values of 
$\fnl$ and $\tnl$ have to be measured in order to either confirm or disprove the SY 
inequality ? 
As we shall see, even though the contributions from $\fnl$ and $\tnl$ are
degenerate in the non-Gaussian halo bias, combining multiple halo mass bins can greatly
help breaking the degeneracy.
As we shall demonstrate, testing  multi-field models of inflation at the 3-$\sigma$ level 
would require, for a {\small EUCLID}-like survey, a detection of a four-point correlator 
amplitude in the collapsed limit of the order of $\tnl\sim 10^5$ given a measurement of a 
local bispectrum at the level of $\fnl\sim 10$. Conversely, we will argue that disproving 
multi-field models of inflation would require a detection of $|\fnl|$ at the level of 80 
or larger if dark matter halos can be resolved down to a mass $10^{10}M_\odot/h$.

The paper is organised as follows. Section 2 contains a short summary of the impact of 
primordial NG on the halo bias at large scales. Section 3 describes the methodology 
adopted. The last Section presents the results and discusses their implications. 
In all illustrations, the cosmology is a flat $\Lambda$CDM Universe with normalisation 
$\sigma_8=0.807$, hubble constant $h_0=0.701$ and matter content $\Omega_{\rm m}=0.279$. 

\section{Non-Gaussian halo bias}
\label{sec:ngbias}

The effect of primordial non-Gaussianity on the halo bias can be computed through various 
methods such as high peaks (\cite{MV08}, \cite{SDH}) or multivariate bias expansions 
(\cite{M08}, \cite{GP10}) but, to date, the peak-background split provides the most 
accurate estimate of the effect (\cite{slosar}, \cite{SK10}, \cite{DJS1}, \cite{SFL12}, 
\cite{SH12}). As shown in \cite{DJS1}, the non-Gaussian contribution to the linear bias is
\begin{align}
\label{eq:dbk}
\Delta b_1(k) &= \frac{4}{(N-1)!} 
\frac{{\cal F}_s^{(N)}\!(k,z)}{{\cal M}_s\!(k,z)} \\
& \times \left[b_{N-2}\delta_c+b_{N-3}
\left(N-3+\frac{d\ln{\cal F}_s^{(N)}\!(k,z)}{d\ln\sigma_s}\right)\right] ~,
\nonumber
\end{align}
where $b_N$ are Lagrangian bias parameters, $\delta_c\sim 1.68$ is the critical threshold
for (spherical) collapse and $\sigma_s$ is the rms variance of the density field at 
redshift $z$ smoothed on the (small) scale $R_s$ of a halo.
The linear matter density contrast $\delta_{\vec{k}}(z)$ is related to the  curvature 
perturbation $\Phi_{\vec{k}}$ during matter domination via the Poisson equation. The latter
can be expressed as $\delta_{\vec{k}}(z)={\cal M}(k,z)\,\Phi_{\vec{k}}$, where
\be
{\cal M}(k,z)\equiv \frac{2}{3}\frac{D(z)}{\Omega_m H_0^2}\,T(k)\,k^2\, . 
\ee
Here, $T(k)$ is the matter transfer function, $\Omega_m$ and $H_0$ are the matter density in 
critical units and the Hubble rate today, and $D(z)$ is the linear growth rate. 
${\cal M}_s$ denotes ${\cal M}(k,z) W(k R_s)$, where $W(k R_s)$ is a spherically symmetric 
window function (we adopt a top-hat filter throughout this paper). 
Furthermore,
\begin{align}
{\cal F}_s^{(N)}\!(k,z) &= \frac{1}{4\sigma_s^2 P_\phi(k)}
\left[\prod_{i=1}^{N-2}\int\!\!\frac{{\rm d}^3k_1}{(2\pi)^3}\,{\cal M}_s(k_i,z)\right]
{\cal M}_s(q,z) \nonumber  \\
& \quad \times \xi_\Phi^{(N)}\!(\vk_1,\cdots,\vk_{N-2},\vq,\vk)
\end{align}
is a projection factor whose $k$-dependence is dictated by the exact shape of the $N$-point 
function $\xi_\Phi^{(N)}$ of the gravitational potential. For the local constant-$\fnl$ model, 
the factor ${\cal F}_s^{(3)}$ is equal to $\fnl$ in the low $k$-limit (squeezed limit), so 
that the logarithmic derivative of ${\cal F}_s^{(N)}$ w.r.t. the rms variance $\sigma_s$ of 
the small-scale density field vanishes on large scales. 
For all other models of primordial non-Gaussianity however, this term is significant for most 
relevant peak heights and becomes negligible in the high peak limit only (\cite{DJS2}). 
Since the halo mass function may not be universal, the non-Gaussian bias correction should 
in principle be computed by taking derivative of the Gaussian halo mass function w.r.t. mass 
(\cite{SH12}). However, because it is difficult to estimate such a mass derivative from real 
data, we will use Eq.(\ref{eq:dbk}), which is valid for a universal mass function. 
Nevertheless, one should bear in mind that non-universality can induce additional corrections 
at the $\sim 10$\% level (\cite{SH12,M12}). Note also that path integral extensions of the 
excursion set formalism (see \cite{MR10}) suggest that memory terms (involving $N$-point 
correlators of the density field smoothed on any scale between $R_s$ and $R_l$) could also 
contribute at some level (\cite{dbk_eps1,dbk_eps2}).  

\begin{table}
\centering
\caption{Average host halo mass, number density, (Lagrangian) linear and quadratic 
bias factors for the low- and high-mass halo samples used in Fig.\ref{fig:ellipse}.}
\begin{tabular}{lcccc}
\toprule
$\mbox{}$& $M (M_\odot/h)$ & $\bar{n}$ ($h^3 {\rm Mpc}^{-3}$) & $b_1$ & $b_2$ \\
\hline\hline
Halo 1 & $10^{12}$ & $7\times 10^{-4}$ & 0.2 & -0.2 \\
Halo 2 & $10^{14}$ & $3\times 10^{-6}$ & 2.5 & 4.5  \\
\bottomrule
\end{tabular}
\label{tab:data}
\end{table}

Specialising the above result to the bispectrum and trispectrum shapes considered here, the 
non-Gaussian bias correction reads
\begin{align}
\label{eq:biasPBS}
\Delta b_1(k) &= 2 \fnl \frac{\delta_c b_1}{{\cal M}_s(k)} +\frac{1}{2}
\left(\gnl+\frac{25}{27}\tnl\right) \\
& \quad\times \frac{\sigma_s^2 S_{s,{\rm loc}}^{(3)}}{{\cal M}_s(k)}
\left[b_2 \delta_c + b_1
\left(1+\frac{d\ln S_{s,{\rm loc}}^{(3)}}{d\ln\sigma_s}\right)\right] \nonumber,
\end{align}
where $b_1$, $b_2$ are the first- and second-order Lagrangian bias parameters, $\gnl$ is 
another NG coefficient parametrising the NG arising from a cubic third-order  term in the 
curvature perturbation $\zeta$ and 
$S_{s,{\rm loc}}^{(3)}(M)$ is the skewness of the density field in a local, quadratic 
non-Gaussian model with $\fnl=1$. 
Strictly speaking, this expression is valid in the limit $k\ll 1$ only since we have 
ignored the $k$-dependence of ${\cal F}_s^{(4)}$. However, deviations become significant 
only for $k\gtrsim 0.1$ where the non-Gaussian signal is negligible and the signal-to-noise
saturates (\cite{SCD11}).
In linear theory, the product $\sigma_s S_{s<{\rm loc}}^{(3)}(M)$ is independent of 
redshift. Therefore, at fixed values of $b_1$ and $b_2$, the non-Gaussian correction 
induced by $\fnl$ scales as $D(z)^{-1}$, whereas that induced by $\gnl$ and $\tnl$ 
does not have any extra dependence on redshift.
For the cosmology considered here, the empirical relation
$\sigma_s S_{s,{\rm loc}}^{(3)}\approx 3.08\times 10^{-4}\sigma_s^{0.145}$ accurately
reproduces the mass dependence of the skewness (\cite{DJS1}). 
The relative amplitude of the $\tnl$-induced scale-dependent bias thus is 
$(\tnl/\fnl)10^{-4}$. Whereas it is negligible in single-field inflation, it can be 
significant for models with $|\tnl|\gg\fnl^2$. Note that current limits from the CMB 
trispectrum are $-0.6<\tnl/10^4<3.3$ (\cite{cmbtnl}).

An important feature of the NG bias correction is that its scale-dependence is degenerate 
in $\fnl$,  $\tnl$ and  $\gnl$ in the large scale limit, since all the $k$-dependence 
is then located in ${\cal M}_s(k) {\sim} 1/k^2$. This degeneracy can be partly broken by 
considering galaxy populations tracing halos of different mass and, possibly, at different
redshifts. 
While recent studies have analysed the problem of detecting NG through future large-scale
surveys combining a number of observational datasets with simple models where only $\fnl$
is nonzero and the other two nonlinear parameters are set to zero, we will assume here that 
both $\fnl$ and $\tnl$ are non-vanishing since we aim at testing the SY inequality 
(\ref{SYin}).
We will however set $\gnl$ to zero
\footnote{Notice that this assumption also gets rid of potentially large one-loop 
corrections to the SY inequality (\cite{by}). These corrections would be anyway below 
the errors we will estimate on  $\tnl$ even for $\gnl$ as large as $10^6$.}. 
We refer the reader to \cite{porciani} for a recent study in which both $\fnl$ and $\gnl$ 
are nonzero. Finally, it is worth mentioning that our Eq.(\ref{eq:biasPBS}) is different 
from the expression given in \cite{GY}, who neglected the mass-dependence of the skewness.

\begin{figure}
\center 
\resizebox{0.40\textwidth}{!}{\includegraphics{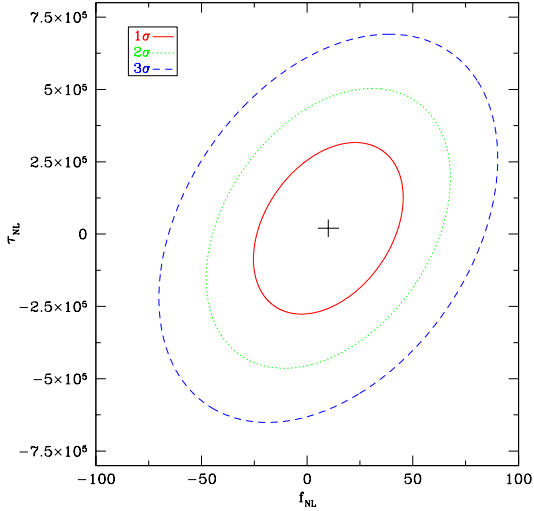}}
\caption{Confidence ellipses obtained by combining the low- and high-mass sample, 
assuming $\fnl=10$ and $\tnl= 2\times 10^4$.}
\label{fig:ellipse}
\end{figure}

\section{Method}
\label{sec:method}

In order to assess the ability of forthcoming experiments to test the SY inequality 
through the measurement of the large scale bias, we use of the Fisher information 
content on $\fnl$ and $\tnl$ from the two-point statistics of halos and dark matter 
in Fourier space. 
The Fisher matrix formalism has been extensively applied to predict how well galaxy
surveys will constrain the nonlinear parameter $\fnl$ (e.g., \cite{dalal}, \cite{CVM}, 
\cite{CHD}). 
In particular, combining differently biased tracers of the same surveyed volume and 
weighting halos by mass can help mitigate the effect of cosmic variance and shot 
noise and, therefore, reduce the uncertainty on $\fnl$ 
(\cite{seljak}, \cite{slosar}, \cite{noise}, \cite{HSD}). 

\subsection{Fisher matrix formalism}
\label{sec:fisher}

Here and henceforth, we closely follow the notation of \cite{HSD} and define the 
halo overdensity in Fourier space as a vector, every element corresponding to halos 
with different mass bins 
\begin{equation}\label{eq:bin}
{\bs \delta}_h = (\delta_h(M_1), \delta_h(M_2), \cdots, \delta_h(M_n))^\top ~.
\end{equation}
Assuming the halos to be locally biased and stochastic tracers of the dark matter 
density field $\delta$, we can write the overdensity of halos as
\begin{equation}
{\bs \delta}_h= {\bf b}\,\delta + {\bs \epsilon} ~,
\end{equation}
where {\bf b} is a vector whose $i$-component is the (Eulerian) bias of the $i$-th
sample,
\begin{equation}
b_i^E(k,M_i,z)= 1 + b_1(M_i,z) + \Delta b_1(k,M_i,z)~,
\end{equation}
and ${\bs \epsilon}$ is a residual noise-field with zero mean. We assume that it is 
uncorrelated with the dark matter. 

Computing the Fisher information requires knowledge of the covariance matrix of the 
halo samples,
\begin{equation}\label{cov}
{\bf C_h} = \langle {\bs \delta}_h {\bs \delta}_h^\top \rangle = 
{\bf b}{\bf b}^\top P + {\bf E}~.
\end{equation}
The brackets indicate the average within a $k$-shell in Fourier space. 
$P=\langle \delta^2 \rangle$ is the non-linear dark matter power spectrum which, on 
large scales, can be assumed independent of $\fnl$ and $\tnl$ and 
${\bf E}=\langle {\bs \epsilon}{\bs \epsilon}^\top\rangle$ is the shot-noise matrix. 
We will follow the general treatment of \cite{HSD} and assume that ${\bf E}$ is not
simply diagonal with entries consistent with Poisson noise (see \S\ref{sec:halomodel}
for explicit expressions).

In order to simultaneously constrain $\fnl$ and $\tnl$, it is pretty clear that at least 
two different halo samples are required to break some of the parameter degeneracies, 
since the bias coefficients $b_1$, $b_2$, the rms variance $\sigma_s$ and the skewness 
$S_{s,{\rm loc}}^{(3)}$ have distinct mass dependences (as is apparent from the numerical
fits of \cite{riotto1} or \cite{enqvist}). More precisely, the Fisher matrix takes the 
following general form
\begin{equation}
\mathcal{F}_{ij} = 
V_{\rm surv}\,f_{\rm sky} \int \frac{{\rm d}k k^2}{2\pi^2}\frac{1}{2} 
{\rm Tr}\,\!\biggr( \frac{\partial {\bf C}_h}{\partial \theta_i}{\bf C}_h^{-1}
\frac{\partial {\bf C}_h}{\partial \theta_j}{\bf C}_h^{-1}\biggr)\, ~,
\label{eq:trace}
\end{equation}
where $\theta_{i}$ are the parameters whose error we wish to forecast. The integral over 
the momenta runs from $k_{\rm min}=\pi/(V_{\rm surv})^{1/3}$ to 
$k_{\rm max}=0.1\, {\rm {\rm Mpc}}^{-1}/ h$, where $V_{\rm surv}$ is the surveyed volume
and $f_{\rm sky}$ is the fraction of the sky observed. For illustration purposes, we will 
adopt the specifications of an {\small EUCLID}-like experiment: $V_{\rm surv}f_{\rm sky}=25$ 
Gpc$^3/h^3$ at median redshift $z=0.7$. We will ignore the redshift evolution 
and assume that all the surveyed volume is at that median redshift. In principle however, 
it should be possible to extract additional information on the non-Gaussian bias from the
redshift dependence of the survey.
For a single mass bin, the four entries of the Fisher matrix have the same $k$-dependence 
at low-$k$. As a consequence, the determinant is very close to zero and, therefore, yields 
large (marginalised) errors. In this case, it is impossible to test the SY inequality 
regardless the characteristics of the halo sample, unless one has some prior on one of the 
parameters. 

\begin{figure*}
\center 
\resizebox{0.40\textwidth}{!}{\includegraphics{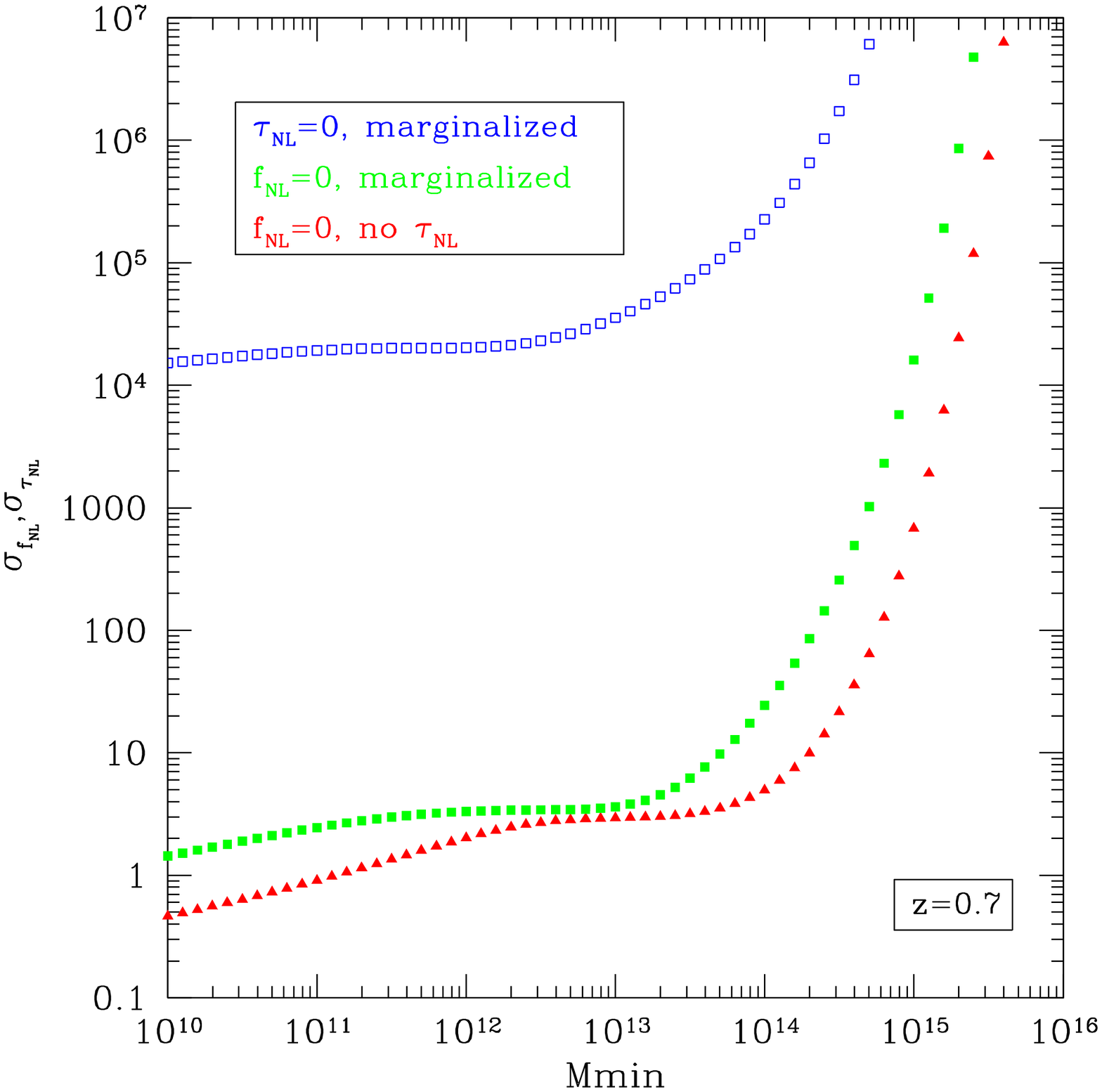}}
\resizebox{0.40\textwidth}{!}{\includegraphics{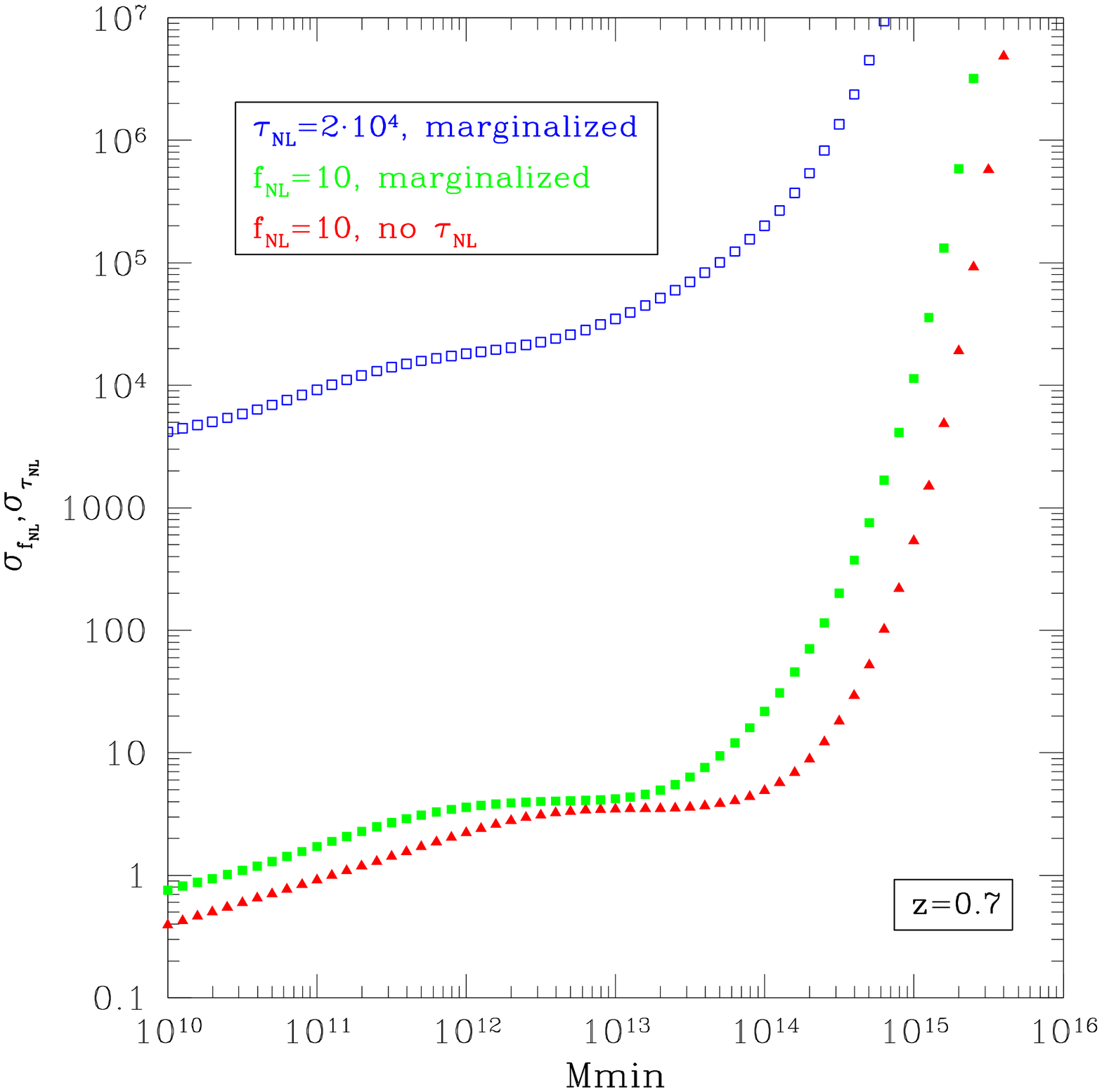}}
\caption{Halo model predictions for the 1-$\sigma$ errors as a function of minimum halo 
mass in the limit of $N\gg 1$ halo mass bins with identical number density. Red triangles
show the uncertainty on $\fnl$ for a one-parameter model with $\fnl=0$ (left panel) and 
$\fnl=10$ (right panel).  Filled (green) and empty (blue) squares represent the 1-$\sigma$
uncertainties on $\fnl$ and $\tnl$ for the two-parameters fiducial models 
$(\fnl,\tnl)=(0,0)$ (left panel) and $(\fnl,\tnl)=(10,2 \times 10^4)$ (right panel).}
\label{fig:mmin}
\end{figure*}

\begin{table}
\centering
\caption{1-$\sigma$ errors obtained with $N\gg 1$ halo mass bins with $M>M_{\rm min}$.
Top and bottom rows show results for $M_{\rm min}=10^{13}$ and $10^{10} M_{\odot}/h$, 
respectively.}
\begin{tabular}{ccccc}
\toprule
$\mbox{}$ & $\fnl=0$ & $\fnl=10$ & $\fnl=0 $ & $ \fnl=10$\\
$\mbox{}$ & no $\tnl$ & no $\tnl$ & $\tnl=0$ & $\tnl=2\times 10^4$\\
\hline\hline
$\sigma_{\fnl}$ & $2.9$ & $3.4$ & $3.6$ & $4.2$\\
$\sigma_{\tnl}$ & $-$ & $-$ & $3.6 \times 10^4$ & $3.4 \times 10^4$\\
\midrule
$\sigma_{\fnl}$ & $0.5$ & $0.4$ & $1.4$ & $0.8$\\
$\sigma_{\tnl}$ & $-$ & $-$ & $1.5 \times 10^4$ & $0.4 \times 10^4$\\
\bottomrule
\end{tabular}
\label{table:table2}
\end{table}

In the general case of $N$ halo populations, the entries of the halo covariance matrix 
read ($a,b = 1,\cdots,N$)
\begin{equation}\label{eq:covdou}
C_{ab}=b_1^E(k, M_a, z) b_1^E(k, M_b, z) P(k) + E_{ab}\, .
\end{equation}
The derivative of the halo covariance matrix with respect to some parameter $\theta$ is
\begin{equation}\label{eq:deriv}
\frac{\partial C_h}{\partial \theta}=({\bf b}_{\theta} {\bf b}^\top 
+ {\bf b}{\bf b}_{\theta}^\top)P\, ,
\end{equation}
where ${\bf b}_{\theta} = \partial {\bf b}/\partial \theta$. We have ignored the dependence 
of ${\bf E}$ on $\theta$ as it is expected to be small for $\fnl$ and $\tnl$. Following 
\cite{HSD}, the inverse of the covariance matrix can be obtained using the Sherman-Morrison 
formula (see \cite{SM} and \cite{bar})
\begin{equation}\label{eq:inverse}
{\bf C}_h^{-1} = {\bf E}^{-1} - 
\frac{{\bf E}^{-1}{\bf b}{\bf b}^\top{\bf E}^{-1}P}{1+{\bf b}^\top{\bf E}^{-1}{\bf b}P}\, .
\end{equation}
On inserting the expression (\ref{eq:deriv}) into Eq. (\ref{eq:trace}), we can write down 
the Fisher matrix for two generic parameters  $\theta_i$ ($i=1,2$) as
\begin{equation}
\mathcal{F}_{ij} = 
\frac{P^2}{2}{\rm Tr}\Bigl[({\bf b}{\bf b}_{i}^\top 
+ {\bf b}_{i}{\bf b}^\top){\bf C}_h^{-1}({\bf b}{\bf b}_{j}^\top
+ {\bf b}_{j}{\bf b}^\top){\bf C}_h^{-1}\Bigr]\, .
\end{equation}
The elements of the Fisher matrix can be easily expressed in terms of the following 
quantities
\begin{align}
\alpha &= {\bf b}^\top{\bf E}^{-1}{\bf b}\, P \\ 
\beta_{i} &= {\bf b}^\top{\bf E}^{-1}{\bf b}_{\theta_i}\, P \nonumber \\
\gamma_{ij} &= {\bf b}_{\theta_i}^\top{\bf E}^{-1}{\bf b}_{\theta_j}\, P 
\nonumber \, .
\end{align}
After some algebra, we obtain 
\begin{equation}
\mathcal{F}_{ij}= \frac{\alpha\gamma_{ij} + \beta_{i}\beta_{j} 
+ \alpha(\alpha\gamma_{ij}-\beta_{i}\beta_{j})}{(1+\alpha)^2}\, ,
\end{equation}
which generalises the calculation reported in \cite{HSD}. Note that, in what follows, 
$\theta_1=\fnl$ and $\theta_2=\tnl$. 

One should bear in mind the caveat that the present Fisher matrix analysis assumes 
Gaussian uncertainties, even though it is likely that the estimators $\hat{f}_{\rm NL}$ 
and $\hat{\tau}_{\rm NL}$ have non-Gaussian distributions. One possible way of testing 
this assumption would be to generate Monte-Carlo simulations of the halo samples, but 
this is beyond the scope of this paper.

\subsection{Halo model predictions}
\label{sec:halomodel}

Even though the halo model makes a number of predictions that are not physically 
sensible (such as a white noise contribution in the limit $k\to 0$ of the cross 
halo-mass power spectrum), it was shown to furnish a very good fit to the eigenvalues 
and eigenvectors of the halo stochasticity matrix (\cite{eigen}).
In this model, the shot-noise matrix can be cast into the closed form expression
${\bf E}$,
\begin{equation}
\label{eq:snmatrix}
{\bf E}=\bar{n}{\bf I}-{\bf b}{\bf\cal M}^\top - {\bf\cal M}{\bf b}^\top ~.
\end{equation}
Here, ${\bf\cal M}={\bf M}/\bar{\rho}_m-{\bf b}\AVE{n{\bf M}^2}/2\bar{\rho}_m^2$, 
${\bf M}$ is a vector whose entries are the halo masses and $n=n(M)$ is the number 
density of halos of mass $M$. The Poisson expectation is recovered upon setting 
${\bf\cal M}=0$. In the limit of $N\gg 1$ halo mass bins with identical number density 
$\bar{n}$, we can replace the scalar products by integrals. 
A straightforward calculation shows that the coefficients $\alpha$, $\beta_i$ and 
$\gamma_{ij}$ can be rewritten as
\begin{align}
\alpha &= \frac{\AVE{b^2}}{\left(\nt^{-1}-\AVE{{\cal M}b}\right)^2
-\AVE{b^2}\AVE{{\cal M}^2}}\nt^{-1} P \\
\beta_i &= \frac{\AVE{b b_{\theta_i}}\left(\nt^{-1}-\AVE{{\cal M}b}\right)+\AVE{b^2}
\AVE{{\cal M}b_{\theta_i}}}{\left(\nt^{-1}-\AVE{{\cal M}b}\right)^2-\AVE{b^2}
\AVE{{\cal M}^2}}P \\
\gamma_{ij} &=
\AVE{b_{\theta_i}b_{\theta_j}}\nt P \\
& \quad +\frac{\AVE{b^2}\AVE{{\cal M}b_{\theta_i}}\AVE{{\cal M}b_{\theta_j}}
+\AVE{{\cal M}^2}\AVE{bb_{\theta_i}}\AVE{bb_{\theta_j}}}
{\left(\nt^{-1}-\AVE{{\cal M}b}\right)^2-\AVE{b^2}\AVE{{\cal M}^2}}\nt P \nonumber \\
& \quad +\frac{\left(\AVE{bb_{\theta_i}}\AVE{{\cal M}b_{\theta_j}}
+\AVE{bb_{\theta_j}}\AVE{{\cal M}b_{\theta_i}}\right)
\left(\nt^{-1}-\AVE{{\cal M}b}\right)}
{\left(\nt^{-1}-\AVE{{\cal M}b}\right)^2-\AVE{b^2}\AVE{{\cal M}^2}}\nt P \nonumber
\end{align}
where 
\begin{align}
\AVE{x y} &\equiv 
\frac{1}{\nt}\int_{M_{\rm min}}^{M_{\rm max}}\!\!dM\,n(M) x(M) y(M) \\
\nt &\equiv \int_{M_{\rm min}}^{M_{\rm max}}\!\!dM\, n(M) = N \bar{n} ~.
\end{align}
Here, $n(M)$ is the halo mass function, which we assume to be of the \cite{ST} form
with $p=0.3$, $q=0.73$ and a normalisation $A=0.322$. 
This yields $\AVE{n{\bf M}^2}/\bar{\rho}_m^2=75.9$Mpc$^3/h^3$ at redshift $z=0.7$.

\section{Results and conclusions} 
\label{sec:results}

We first compute the uncertainties on $\fnl$ and $\tnl$ from two different tracer 
populations and for a shot-noise matrix consistent with Poisson noise, i.e.
${\bf E}={\rm diag}(1/n(M_1),1/n(M_2))$.
We consider a nearly unbiased sample with average mass $M\sim 10^{12}M_\odot/h$ 
and a high mass sample with $M=10^{14}M_\odot/h$. Table \ref{tab:data} summaries 
the characteristics of these populations. For a given mass $M$, the second-order 
Lagrangian bias parameter $b_2(M)$ is computed from the Sheth-Tormen multiplicity 
function, whereas the skewness $S_{s,{\rm loc}}^{(3)}$ is computed from the 
phenomenological relation given in \S\ref{sec:ngbias}. 
Fig.\ref{fig:ellipse} shows the resulting 68, 95 and 99\% confidence contours for 
the parameters $\fnl$ and $\tnl$ when the fiducial model assumes $\fnl=\pm 10$ and 
$\tnl = 2\times 10^4$. The 1-$\sigma$ errors are $\sigma_{\fnl}\simeq 23$ and 
$\sigma_{\tnl}\simeq 2.0\times 10^5$.
We have tried different combinations of halo populations and found that the 
errors do not change significantly. At this point, we would conclude that galaxy
bias alone cannot yield interesting constraints on $\tnl$ and $\fnl$.

The situation changes dramatically when the surveyed halos are divided into $N\gg 1$ 
populations of increasing mass, with equal number density. 
In Fig. \ref{fig:mmin}, symbols represent the halo model prediction for the 1-$\sigma$
uncertainties $\sigma_{\fnl}$ and $\sigma_{\tnl}$ in the limit of infinitely many halo 
bins. The shot-noise matrix now takes the form Eq.(\ref{eq:snmatrix}).
Red triangles indicate $\sigma_{\fnl}$ in a one-parameter model with $\fnl=0$ 
(left panel) and $\fnl=10$ (right panel). Filled and empty squares represent 
$\sigma_{\fnl}$ and $\sigma_{\tnl}$ in a two-parameters model with $(\fnl,\tnl)=(0,0)$ 
(left panel) and $(\fnl,\tnl)=(10,2\times 10^4)$ (right panel).  
Results are shown as a function of the mass of the smallest halos resolved in the 
survey. Compared to the previous configuration, significant gains are already achieved
for $M_{\rm min}\approx 10^{13}M_\odot/h$. While the constraint on $\fnl$ is somewhat 
degraded if one allows for a non-zero $\tnl$, the 1-$\sigma$ uncertainty on $\tnl$ is 
of the order of $(10^3 - 10^4)$, an order of magnitude better than in the case of two 
galaxy populations. Table \ref{table:table2} gives the 1-$\sigma$ errors for 
$M_{\rm min}=10^{13}$ and $10^{10}M_\odot/h$.

\begin{figure}
\center 
\resizebox{0.40\textwidth}{!}{\includegraphics{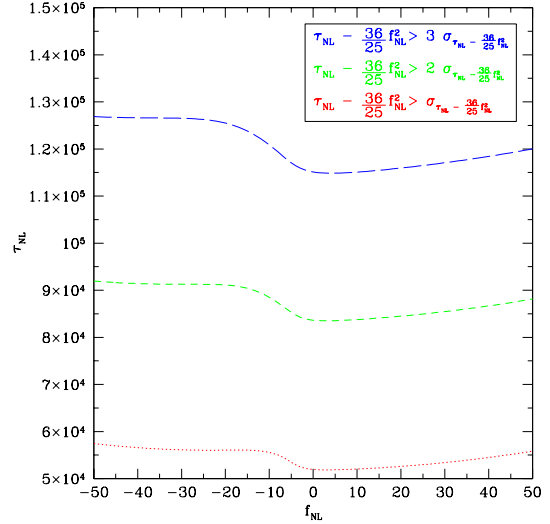}}
\caption{Testing the validity of the SY inequality with measurements of the 
non-Gaussian bias. The various curves are halo model predictions for $N\gg 1$
halo mass bins with $M_{\rm min}=10^{13}M_\odot/h$. At fixed value of $\fnl$, 
they indicate the minimum $\tnl$ required in order to have a measurement of 
the SY inequality at the 1-, 2- and 3-$\sigma$ confidence level.}
\label{fig:set1}
\end{figure}

How well can we test the SY inequality with galaxy bias?
Fig. \ref{fig:set1} displays, as a function of $\fnl$,  the minimum value of $\tnl$ for 
which the difference $(\tnl-(36/25)\fnl^2)$ is greater than its 1-,2- and 3-$\sigma$ 
error which, for Gaussian-distributed $\fnl$ and $\tnl$, reads
\begin{align}
\sigma^2_{\tnl-\frac{36}{25}\fnl^2} &= 
\sigma^2_{\tnl}+2\left(\frac{36}{25}\right)^2\sigma_{\fnl,\tnl}^2 \\
& \quad + 4\left(\frac{36}{25}\right)^2\sigma^2_{\fnl}\bar{f}^2_{\rm NL}
-4\left(\frac{36}{25}\right)\sigma_{\fnl,\tnl}\bar{f}_{\rm NL} \nonumber \;,
\end{align}
where $\sigma^2_{\fnl}, \sigma_{\fnl,\tnl}$ and $\sigma^2_{\tnl}$ are the entries of 
the inverted Fisher matrix and $\bar{f}_{\rm NL}$, $\bar{\tau}_{\rm NL}$ are the values of 
the fiducial model assumed.
The various curves indicate the halo model prediction for $N\gg 1$ halo populations 
with a minimum resolved mass $M_{\rm min}=10^{13}M_\odot/h$.
For instance, if a non-vanishing value of $\fnl=10$ is measured in the future, then 
the contribution induced by the collapsed limit of the trispectrum must be detected 
with an amplitude of at least $\tnl \sim {\cal O}(1)\times 10^5$ in order to have a 
3-$\sigma$ detection of the SY inequality with the non-Gaussian halo bias.
Of course, these values are only indicative since the analysis is performed with the 
restrictive assumption of Gaussian errors.

Finally, we can also assess how well galaxy bias can probe the violation of the SY 
inequality. As stated above, the observation of a strong  violation would have profound 
implications for inflationary models as it implies either that multi-field inflation, 
independently of the details of the model, cannot be responsible for generating the 
observed fluctuations, or that some new non-trivial (ghost-like) degrees of freedom 
play a role during inflation. 
Measuring a violation essentially consists in a simultaneous detection of  a non-zero 
value of $\fnl$ and a (non-zero) small enough value of $\tnl$. Here, we have simply 
estimated the smallest $|\fnl|$ such that $(36/25)\fnl^2$ is larger than the 3-$\sigma$
error on $\tnl$. Having found that, for the current observationally allowed range of 
$\fnl$, the error of $\tnl$ does not significantly change if we set in all runs $\tnl=0$, 
we have thus computed $\sigma_{\tnl}$ assuming a vanishing value of $\tnl$. 
A comparison of $3\sigma_{\tnl}$ with $(36/25)\fnl^2$ shows that, for a minimum halo mass
$M_{\rm min}=10^{13}M_\odot/h$, the SY inequality cannot be tested with the non-Gaussian 
galaxy bias solely for realistic values of $\fnl$. Even if halos are resolved down to 
$10^{10}M_\odot/h$ is $3\sigma_{\tnl} < (36/25)\fnl^2$ satisfied only for $|\fnl|$ larger 
than $\sim$ 80. 

Summarising,  a large NG in the squeezed limit implies that the cosmological perturbations 
are generated by some light scalar field other than the inflaton. The  SY inequality 
(\ref{SYin}) inevitably imposes that a large trispectrum in the collapsed limit is also 
present. 
However, the contribution of $\tnl$ to the non-Gaussian halo bias is suppressed by 
$10^{-4}(\tnl/\fnl)$ and strongly degenerate with that induced by $\fnl$.
Notwithstanding this, we have shown that multi-tracer methods can exploit the distinct 
mass-dependence of the $\fnl$- and $\tnl$-induced bias corrections to reduce the 1-$\sigma$ 
uncertainty down to $\sigma_{\tnl}\lesssim 10^4$ (and simultaneously achieve 
$\sigma_{fnl}\sim 1-5$) for a survey covering half of the sky up to $z\approx 1$. 
The exact values depend on the mass $M_{\rm min}$ of the least massive halos observed. 
Our results on the capability of testing  the SY inequality through the NG scale-dependent 
bias are summarised in Fig.\ref{fig:set1}. The latter shows that testing the SY inequality 
at the level of 3-$\sigma$ would require detecting $\tnl$ at the level of $10^5$ for the 
minimum resolved mass  $M_{\rm min}=10^{13}M_\odot/h$. 
Conversely, testing the violation of the SY inequality requires both a much smaller 
resolved mass, $M_{\rm min}=10^{10}M_\odot/h$ and a large bispectrum, $|\fnl|\gtrsim 80$. 
As mentioned above, all these results are valid provided that $\gnl=0$ and that the 
nonlinear parameters $\fnl$ and $\tnl$ estimated from the data are Gaussian-distributed. 
Relaxing these assumptions will be the subject of future work.

\section*{Acknowledgements}

We thank Licia Verde for comments on an early version of this manuscript, and  
Uro\v s Seljak for discussions.
M.B. and V.D. acknowledge support by the Swiss National Science Foundation. 
A.R. is supported by the Swiss National Science Foundation, project `The 
non-Gaussian Universe" (project number: 200021140236).

\bibliographystyle{mn2e}

\begin{thebibliography}{99}


\bibitem[Acquaviva  {et al.} (2003)]{acqua}
Acquaviva V., Bartolo N., Matarrese S.  \& Riotto A., 2003,
Nucl.\ Phys.\ B {\bf 667}, 119.

\bibitem[Ashead {et al.} (2012)]{dbk_eps2}
Ashead P., Baxter E.J., Dodelson S. \& Lidz A., 2012,
arXiv:1206.3306.

\bibitem[Assassi {et al.} (2012)]{SY3}
Assassi V., Baumann D.  \& Green D., 2012,
arXiv:1204.4207 [hep-th].

\bibitem[Bartlett (1951)]{bar}
Bartlett M.S., 1951,
Ann. Math. Stat. {\bf 22}, 107.

\bibitem[Bartolo  {et al.}(2004)]{BKMR}
Bartolo N., Komatsu E., Matarrese S. \& Riotto A.,   2004,    
 Phys.\ Rept.\  {\bf 402}, 103.

\bibitem[Bartolo  \& Riotto (2009)]{recomb}
Bartolo N. \& Riotto A.,   2009,    
 JCAP {\bf 0903}, 017.

\bibitem[Tasinato    {et al.}(2012) ]{by}
Tasinato G., Byrnes C., Nurmi S. \& Wands D.,     2012,    
  arXiv:1207.1772 [hep-th].


\bibitem[Carbone {et al.} (2008)]{CVM}
Carbone C., Verde L. \& Matarrese S., 2008,
Astrophys.\ J. {\bf 684}, L1.

\bibitem[Cheung {et al.} (2008)]{con3}
Cheung C., Fitzpatrick L., Kaplan J. \& Senatore L.,   2008,
JCAP {\bf 0802}, 021.

\bibitem[Chongchitnan and Silk (2010)]{CS}
Chongchitnan S. \& Silk J., 2010,
Astrophys.\ J. {\bf 724}, 285.

\bibitem[Creminelli and Zaldarriaga (2004)]{con2}
Creminelli P. \& Zaldarriaga M.,   2004,
JCAP {\bf 0410}, 006.

\bibitem[Cunha {et al.} (2010)]{CHD}
Cunha C., Huterer D. \& Dor\'e O., 2010,
Phys. \ Rev. \ D {\bf 82}, 023004.

\bibitem[Dalal  {et al.} (2008)]{dalal}
Dalal N., Dore O., Huterer D. \& Shirokov A., 2008,
Phys.\ Rev.\ D {\bf 77}, 123514.

\bibitem[D'Aloisio {et al.} (2012)]{dbk_eps1}
D'Aloisio A., Zhang J., Jeong D. \& Shapiro P.R., 2012,
arXiv:1206.3305.

\bibitem[De Simone {et al.} (2011)]{riotto1}
De Simone A., Maggiore M.  \& Riotto A., 2011,
Mon.\ Not.\ Roy.\ Astron.\ Soc.\  {\bf 412},  2587.

\bibitem[Desjacques and Seljak (2010)]{DS10}
Desjacques V. \& Seljak U., 2010,
Phys.\ Rev.\ D {\bf 81}, 023006.

\bibitem[Desjacques  {et al.} (2011a)]{DJS1}
Desjacques V., Jeong D.\& Schmidt F., 2011a,
Phys.\ Rev.\ D {\bf 84}, 063512.

\bibitem[Desjacques  {et al.} (2011b)]{DJS2}
Desjacques V., Jeong D. \& Schmidt F., 2011b,
Phys.\ Rev.\ D {\bf 84}, 061301.

\bibitem[Dvali and Gruzinov (2004)]{rate1}
Dvali G.   \&  Gruzinov A.,  2004,    
 Phys.\ Rev.\ D {\bf 69}, 023505.

\bibitem[Enqvist and Sloth (2002)]{curvaton1}
Enqvist K.   \&  Sloth M.,  2002,    
Nucl.\ Phys.\ B {\bf 626}, 395.

\bibitem[Enqvist {et al.} (2011)]{enqvist}
Enqvist K., Hotchkiss S. \& Taanila O., 2011,
JCAP 1104,  017.

\bibitem[Giannantonio and Porciani (2008)]{GP10}
Giannantonio T. and Porciani C., 2010,
Phys.\ Rev.\ D {\bf 81}, 063530.

\bibitem[Gong and Yokoyama (2011)]{GY}
Gong J.-O. and Yokoyama S., 2011,
Mon.\ Not.\ Roy.\ Astron.\ Soc.\  {\bf 417}, L79.

\bibitem[Hamaus {et al.} (2010)]{eigen}
Hamaus N., Seljak U., Desjacques V., Smith R.E., Baldauf T., 2010,
Phys. \ Rev. \ D {\bf 82}, 043515.

\bibitem[Hamaus {et al.} (2011)]{HSD}
Hamaus N., Seljak U. \& Desjacques V., 2009,
Phys. \ Rev. \ D {\bf 84}, 083509.

\bibitem[Kehagias and Riotto(2012)]{KR}
Kehagias A. \& Riotto A., 2012,
 arXiv:1205.1523 [hep-th], to be published in Nucl. 
 Phys. {\bf B}.

\bibitem[Kofman (2003)]{rate2}
Kofman L.,  2004,  
arXiv:astro-ph/0303614.

\bibitem[Kolb {et al.}(2005)]{during}
Kolb E.W., Riotto A. \& Vallinotto A.,  2005,    
Nucl.\ Phys.\ B {\bf 626}.

\bibitem[Komatsu and Spergel (2001)]{koma}
Komatsu R. \& Spergel D.N., 2001,
Phys. \ Rev. \ D {\bf 63}, 063002.

\bibitem[Komatsu {et al.}(2011)]{wmap7}
Komatsu E.  {\it et al.}  [WMAP Collaboration], 2011
  Astrophys.\ J.\ Suppl.\  {\bf 192}, 18.

\bibitem[Lyth and Riotto(1999)]{lrreview}
Lyth D.   \&  Riotto A.,  1999,    
 Phys.\ Rept.\  {\bf 314}, 1.

\bibitem[Lyth and Wands (2002)]{LW}
Lyth D.   \&  Wands D.,  2002,    
Phys.\ Lett.\ B {\bf 524}, 5.

\bibitem[Lyth (2005)]{end1}
Lyth D.,  2005,  
JCAP {\bf 0511}, 006.

\bibitem[Lyth and Riotto (2006)]{end2}
Lyth D. \& Riotto A.,  2006,  
Phys.\ Rev.\ Lett.\  {\bf 97}, 12130.

\bibitem[Maggiore and Riotto (2010)]{MR10}
Maggiore M. \& Riotto A., 2010,
Astrophys.\ J. {\bf 711}, 907.

\bibitem[Maldacena(2003)]{con1}
Maldacena J.,   2003,
JHEP {\bf 0305}, 013.

\bibitem[Matarrese and Verde (2008)]{MV08}
Matarrese S. \& Verde L., 2008,
Astrophys.\ J. {\bf 677}, L77.

\bibitem[Matsubara (2012)]{M12}
Matsubara T., 2012,
arXiv:1206.0562.

\bibitem[McDonald (2008)]{M08}
McDonald P., 2008,
Phys.\ Rev.\ D {\bf 78}, 123519.

\bibitem[Moroi and Takahashi (2002)]{curvaton3}
Moroi T.   \&  Takahashi T.,  2002,    
Phys.\ Lett.\ B {\bf 522}, 215
[Erratum-ibid.\ B {\bf 539}, 303 (2002)].

\bibitem[Roth and Porciani  (2012)]{porciani}
Roth N. \& Porciani C.~C., 2012,
arXiv:1205.3165.

\bibitem[Schmidt and Kamionkowski (2010)]{SK10}
Schmidt F. \& Kamionkowski M., 2010,
Phys.\ Rev.\ D {\bf 82}, 103002.

\bibitem[Scoccimarro {et al.} (2012)]{SH12}
Scoccimarro R., Hui L., Manera M. \& Chan K.~C., 2012,
Phys.\ Rev.\ D {\bf 85}, 083002.

\bibitem[Sefusatti {et al.} (2011)]{SCD11}
Sefusatti E., Crocce M. \& Desjacques V., 2011,
arXiv:1111.6966.

\bibitem[Seljak (2009)]{seljak}
Seljak U.,  2009,  
Phys. \ Rev. \ Lett. {\bf 102}, 021302.

\bibitem[Seljak {et al.} (2009)]{noise}
Seljak U., Hamaus N. \& Desjacques V., 2009,  
Phys. \ Rev. \ Lett. {\bf 103}, 091303.

\bibitem[Shandera {et al.} (2011)]{SDH}
Shandera S., Dalal N. \& Huterer D., 2011,
JCAP {\bf 1103}, 017.

\bibitem[Sherman and Morrison (1950)]{SM}
Sherman J. \& Morrison W.J., 1950,
Ann. Math. Stat. {\bf 21}, 124.

\bibitem[Sheth and Tormen (1999)]{ST}
Sheth R.K. \& Tormen G., 1999,
Mon.\ Not.\ Roy.\ Astron.\ Soc.\  {\bf 308}, 119.

\bibitem[Slosar {et al.} (2008)]{slosar}
Slosar A., Hirata C.M., Seljak U., Ho S.\& Padmanabhan N., 2008,
JCAP {\bf 0808}, 031.

\bibitem[Slosar (2009)]{slosar}
Slosar A.,  2009,  
JCAP {\bf 0903}, 004.

\bibitem[Smidt {et al.} (2010)]{cmbtnl}
Smidt J., Amblard A., Byrnes C.~T., Cooray A., Heavens A. \& Munshi D., 2010,
Phys.\ Rev.\ D {\bf 81}, 123007.

\bibitem[Smith  {et al.} (2011)]{SY2}
Smith K.M., Lo Verde M. \& Zaldarriaga M., 2011,
Phys.\ Rev.\ Lett.\  {\bf 107}, 191301.

\bibitem[Smith {et al.} (2012)]{SFL12}
Smith K.~M., Ferraro S. \& LoVerde M., 2012,
JCAP {\bf 0312}, 032.
  
\bibitem[Sugiyama {et al.} (2011)]{SY1}
Sugiyama N.S., Komatsu E. \& Futamase T., 2011,
Phys.\ Rev.\ Lett.\  {\bf 106}, 251301.

\bibitem[Suyama and Yamaguchi (2008)]{SY}
Suyama T. \& Yamaguchi M.,   2008,
 Phys.\ Rev.\ D {\bf 77}, 023505. 

\bibitem[Xia {et al.} (2011)]{X11}
Xia J.-Q., Baccigalupi C., Matarrese S., Verde L. and Viel M., 2011,
JCAP{\bf 0811}, 033.

\end{thebibliography}

\label{lastpage}

\end{document}